\documentclass{article}

\usepackage{arxiv}

\usepackage[utf8]{inputenc} 
\usepackage[T1]{fontenc}    
\usepackage{hyperref}       
\usepackage{url}            
\usepackage{booktabs}       
\usepackage{amsfonts}       
\usepackage{nicefrac}       
\usepackage{microtype}      
\usepackage{lipsum}

\usepackage{graphicx}
\usepackage{amsmath,amssymb,amsfonts}
\usepackage{algorithmic}
\usepackage{algorithm}
\usepackage{subcaption}
\usepackage{color}

\title{Data-driven Outer-Loop Control Using Deep Reinforcement Learning for Trajectory Tracking}

\author{
  Maria Angelica.~Arroyo\thanks{Software Engineer at Lockheed Martin, Sikorsky Aircraft Corporation, Stratford, CT 06614.} \\
  Department of Electrical and Electronic Engineering\\
  University of los Andes\\
  Bogota, Colombia 111711 \\
  \texttt{ma.arroyo10@uniandes.edu.co} \\
   \And
 Luis Felipe Giraldo \\
  Department of Electrical and Electronic Engineering\\
  University of los Andes\\
  Bogota, Colombia 111711 \\
  \texttt{lf.giraldo404@uniandes.edu.co} \\
}

\begin{document}
\maketitle

\begin{abstract}
Reference tracking systems involve a plant that is stabilized by a local feedback controller and a command center that indicates the reference set-point the plant should follow. Typically, these systems are subject to limitations such as disturbances, systems delays, constraints, uncertainties, underperforming controllers, and unmodeled parameters that do not allow them to achieve the desired performance. In situations where it is not possible to redesign the inner-loop system, it is usual to incorporate an outer-loop control that instructs the system to follow a modified reference path such that the resultant path is close to the ideal one. Typically, strategies to design the outer-loop control need to know a model of the system, which can be an unfeasible task. In this paper, we propose a framework based on deep reinforcement learning that can learn a policy to generate a modified reference that improves the system's performance in a non-invasive and model-free fashion. To illustrate the effectiveness of our approach, we present two challenging cases in engineering: a flight control with a pilot model that includes human reaction delays, and a mean-field control problem for a massive number of space-heating devices. The proposed strategy successfully designs a reference signal that works even in situations that were not seen during the learning process.
\end{abstract}


\section{Introduction}
\label{sec:introduction}
\begin{figure}[ht]
\begin{center}
\includegraphics[width=0.5\textwidth]{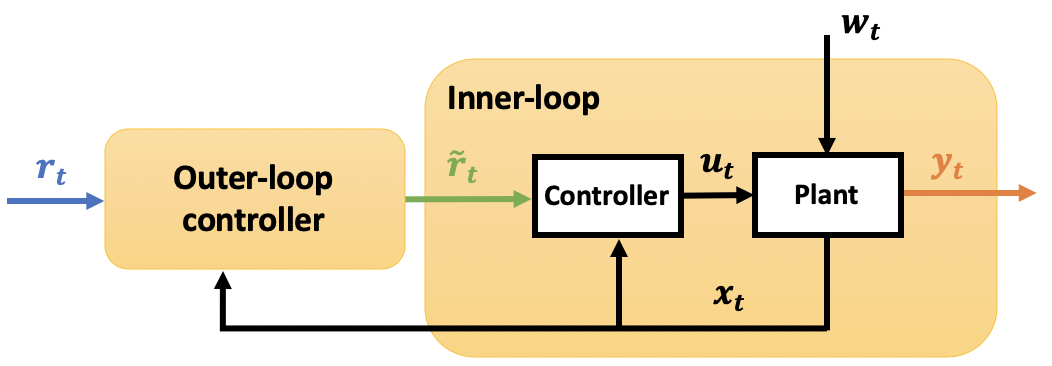}    
\caption{Outer-loop control applied to a inner-loop system.} 
\label{fig:tracking}
\end{center}
\end{figure}

The main objective in trajectory tracking control is to find the appropriate strategy to make the output of a dynamical system follow a pre-defined trajectory. Even though control design strategies such as robust control~\cite{sun}, adaptive control~\cite{lungu}, and model predictive control~\cite{klauco} have effectively achieved high performance and stability in this task, scenarios commonly seen in engineering with implementation issues and constraints, unmeasured disturbances, and uncertainties can affect the control action and deteriorate the system's performance. In situations where it is not possible to redesign or modify the inner-loop, it is feasible to incorporate an outer-loop that commands the inner-loop by adjusting its set point, following the scheme shown in Fig. \ref{fig:tracking}

The main idea in this outer-loop control scheme is to provide a modified reference when the original reference may lead the system to undesired trajectories. This scheme has been studied and implemented for different types of systems and scenarios including linear, nonlinear, and networked and multi-agent systems \cite{cao,naldi,garone,pezzutto2019reference}.   Typically, in such approaches, an important assumption that needs to be considered is that the model of the local controller, plant, and boundaries of the uncertainties have to be known a priori. This assumption is often a complicated or impossible task to conduct. Very little has been done to propose data-driven outer-loop strategies in a trajectory tracking framework for systems where, besides the infeasibility to change its local controller, there is not an accurate model available that characterizes its dynamics \cite{liu2019model,lanchares2019reference}. We address this problem in this paper. 

We hypothesize that Reinforcement Learning (RL) can offer powerful tools for designing a outer-loop control to enforce trajectory tracking in an online and model-free scheme. RL is a branch of artificial intelligence where an agent learns an optimal control policy by maximizing a reward function while it is interacting with the environment. The incorporation of deep neural networks (DNN) as function approximators into the RL framework is known as deep reinforcement learning (DRL) . Complex control problems with discrete and continuous action spaces such as robotic manipulation, bipedal locomotion, and tracking control problems have been successfully solved using DRL~\cite{garcia,aragon,Carlucho,lekkas,Guay}. Inspired by the ideas in RL, in this paper we explore the implementation of a non-invasive, full-online and model-free outer-loop control using DRL in the context of tracking control problems. As far as we know, this has not been done before. We assume that i) the model of the system is unknown, ii) the local controller cannot be modified, and iii) the control signal given by the local controller cannot be observed. The goal of our DRL agent is to improve the performance of the system only using data information of the output variable and the desired reference. The learned optimal policy is used to generate the reference input to the inner-loop system. In a model-free fashion, this modified reference takes into account the uncertainties, imperfections and constraints of this system to achieve the desired performance. 

To show the effectiveness of our proposed data-driven strategy to design outer-loop systems, we present two complex engineering problems: control of human in the loop systems, and control of massive devices in mean field games. The first study case is a challenging problem due to the human reaction delays as a critical parameter for instabilities. We present numerical simulations for an aircraft pitch angle control with the Neal-Schmidt pilot model, that is a first order lead-lag-type compensator with a gain, time lag with time-delay. The second study case involves the control of a very large population of devices using mean field games. We present numerical simulations where the mean temperature of a system composed of a massive number of space heating devices follows a target temperature trajectory using a strategy based on mean field games. Each household is modeled using a stochastic differential equation and has a local linear control law with a poor performance that only requires the information of the target trajectory. We are able to achieve the desired performance with our proposed DRL-based trajectory tracking strategy in a model-free fashion while preserving the local controller of each device in the population.

This paper is organized as follows. In Section \ref{sec:S2} we present our data-driven outer-loop trajectory tracking architecture and the explanation of the DRL elements in a control system context. In Section \ref{sec:HIL}, the human-in-the-loop control problem is introduced and numerical simulations are shown. The case of study mean field games with numerical simulations is presented in Section \ref{sec:MFC}. Finally, in Section \ref{sec:conclusion} conclusions as well as future research directions are given.

\begin{figure*}
\begin{center}
\includegraphics[width=\textwidth]{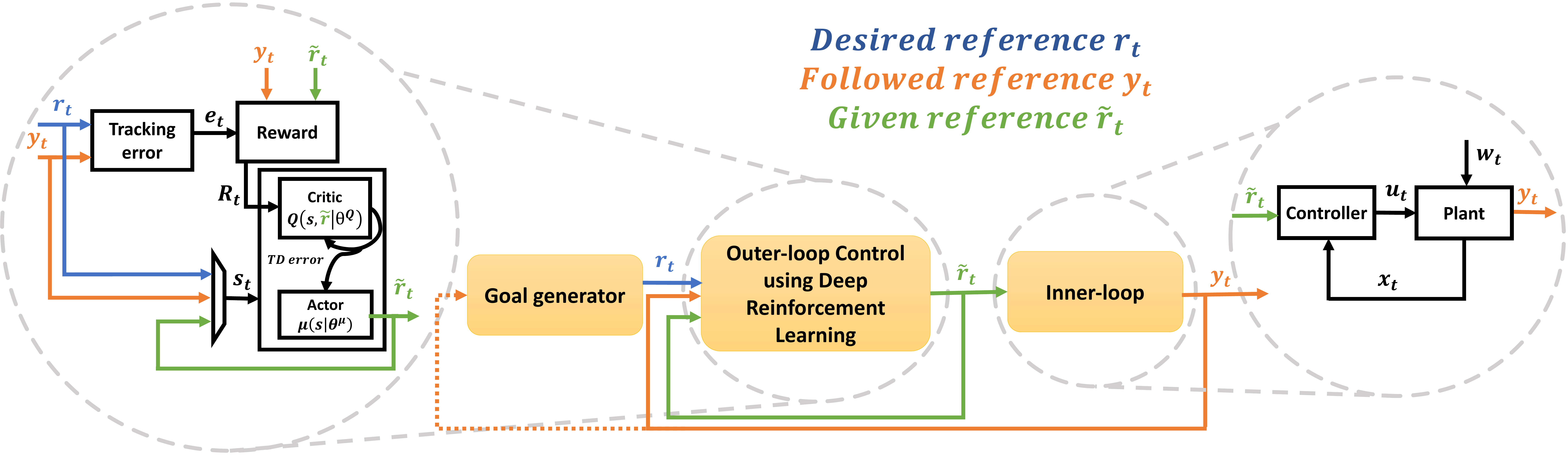}    
\caption{Trajectory tracking scheme applied to a inner-loop system with constraints.} 
\label{bg}
\end{center}
\end{figure*}

\section{Outer-Loop System Design using Deep Reinforcement Learning} \label{sec:S2}
In our DRL control framework, the environment is the controlled dynamical system that takes the action as an input and returns its next reference state. The solution architecture of the data-driven outer-loop control using DRL for trajectory tracking is shown in Fig. \ref{bg}. In this scheme, the command center contains the goal generator and the outer-loop control. The goal generator block indicates the desired reference $r_t$ to the outer-loop control at each time $t$. The main function of this block is to generate a pre-established trajectory in space to be followed by the controlled system. The reference $r_t$ can be feedback controlled by the behavior of the system with $y_t$ as an input. For example, the case of a human operator that interacts with the observed state of the system to indicate the next goal using feedback driving devices such as joystick, flight stick, wheels and touch screens. In the DRL-based outer-loop strategy, we assume that the model of the system is unknown, the local controller cannot be modified, and the control signal given by the local controller cannot be observed. Also, we assume that the local controller leads to a poor performance as a consequence of unmeasured disturbances, parametric uncertainties, model approximations and implementation constraints. The DRL agent learns an optimal policy that determines the modified reference $\widetilde{r}_t$ that will improve the the inner-loop system's performance in an online and model-free fashion. The actual followed trajectory $y_t$ corresponds to the output variables of the inner-loop system.

The reward function is a fundamental component in a DRL framework that measures the success or failure of the agent’s actions in a given state, driving the learning process. In our framework to design a DRL-based outer-loop control, the reward function typically evaluates the ability of the agent to lead the system to the reference trajectory and to satisfy constraints on the inputs and observed outputs. Other important elements are the policy and the action-value function. The policy $\mu$ is the strategy that the agent employs to determine the next action based on the current state. It maps states to actions that maximize the reward. The action-value function $Q_\mu(s_t,\widetilde{r}_t)$ refers to the long-term return of taking action $\widetilde{r}_t$ under policy $\mu$ from the current state $s_t$. Methods to implement RL can be classified in value-based algorithms, policy-based algorithms, and actor-critic algorithms~\cite{lopes}. Value-based algorithms use the action-value function to drive the learning process to obtain the policy. A main disadvantage of this type of algorithms is the need of a discretization of the action space. By contrast, the policy-based algorithms do not use a stored value function. Instead, they work with a parameterized family of policies and optimize the cost function such that the expected value of the sum of rewards is maximized. These algorithms use the gradient of the cost function to conduct the optimization process. A main drawback of this type of algorithms is that the estimated gradient may have a large variance and every gradient is calculated without using knowledge of past estimates~\cite{lopes}. On the other hand, we have actor-critic algorithms, which are a hybrid between value-based and policy-based algorithms. An actor-critic algorithm incorporates two function approximators: the actor and the critic. The role of the actor is to approximate the optimal control policy $\mu$, and the role of the critic is to evaluate the current policy prescribed by the actor while updating and approximating the value function $Q_\mu(s_t,\widetilde{r}_t)$. This value function is used to update the actor's policy parameters for performance improvement. 
In our cases of study, we use deep deterministic policy gradients (DDPG)~\cite{Lillicrap} as our DRL algorithm. DDPG is a popular model-free actor-critic algorithm that estimates a deterministic policy and that can operate over continuous action spaces. Even though we chose to use DDPG, other kind of reinforcement learning algorithms such as Proximal Policy Optimization (PPO)~\cite{PPO} and Trust Region Policy Optimization (TRPO)~\cite{TRPO} can be used in our proposed scheme to design a outer-loop control for trajectory tracking.

Next, we will first propose a general form of a reward function to drive the learning process of the outer-loop control. This reward  will lead the system to minimize the tracking error and to penalize state and reference values that do not satisfy previously defined constraints on each problem. Then, we will explain the DDPG algorithm used to implement the DRL-based outer-loop control.

\subsection{Reward Function}
A critical part of the design of our DRL-based outer-loop control is the definition of the reward function, since the learning process is based on its maximization. We propose an approach where the reward is given by a function that leads the system to minimize the tracking error and to try to satisfy output constraints. Let $e_t = \big|\big|r_t - y_t\big|\big|$ be the absolute tracking error. The proposed reward function is given by
\begin{equation}\label{ec:reward_general}
    R(e_t, y_t, \widetilde{r}_t) = G(e_t)+H(\widetilde{r}_t)+B(y_t)  ,
\end{equation}
where $G$ is a decreasing function that leads the learning process to an error minimization, and $H$ and $B$ are functions that penalize state values and modified reference values (respectively) that do not satisfy constraints that have been previously defined. Functions $G$, $B$, and $H$ are designed depending on the application where the outer-loop scheme is used.

\subsection{Deep Deterministic Policy Gradient Algorithm}
DDPG is a model-free actor-critic algorithm that estimates a deterministic policy and that operates over continuous actions spaces~\cite{Lillicrap}. DDPG uses four parameterized deep neural networks. The first network, called the actor, specifies the current policy by deterministically mapping states $s_t$ to actions $\widetilde{r}_t$ as $\widetilde{r}_t = \mu(s_t|\theta^{\mu})$. The second network, called the critic $Q(s_t,\widetilde{r}_t|\theta^{Q})$, maps $s_t$ and $\widetilde{r}_t$, given by the actor, into the value function that evaluates the long term return (or future rewards) of computing $\widetilde{r}_t$ in $s_t$. Since the policy is deterministic, we allow for exploration during training using a random process $\mathcal{N}$ (called Ornstein-Uhlenbeck process) to add noise to $\widetilde{r}_t$. To minimize temporal correlations between samples, DDPG uses a replay buffer, which stores experiences as $(s_t, \widetilde{r}_t, R_t, s_{t+1})$. When the replay buffer is full, then the oldest experiences are discarded. At each time step, the actor and critic are updated by sampling a minibatch $\mathcal{B}$ of $\mathcal{N}$ from the buffer, allowing for a faster learning than using all experiences.

The other two neural networks in DDPG, known as target networks, are used to regularize values of the actor and critic networks as $\mu^{'}(s_t|\theta^{\mu^{'}})$ and $Q^{'}(s_t,\widetilde{r}_t|\theta^{Q^{'}})$ to increase stability of the learning process. The update rules for the target actor and critic networks are given by
\begin{eqnarray}
        \label{ec:target}
        \theta^{Q'} \leftarrow \tau \theta^{Q}+(1-\tau)\theta^{Q'} \\
        \theta^{\mu'}\leftarrow \tau \theta^{\mu}+(1-\tau)\theta^{\mu'},
\end{eqnarray}
with $\tau\in[0,1]$ being a regularization term. The critic network is updated with the following gradient-based rule 
\begin{eqnarray}
        \label{ec:critic}
        \theta^{Q}&\leftarrow& \theta^{Q}-\alpha_{c} \frac{1}{N}\sum_{i \in \mathcal{B}} \nabla_{\theta^{Q}} (g_{i}-Q(s_i,\widetilde{r}_i|\theta^{Q}))^{2},
\end{eqnarray}
where the learning rate is $\alpha_c$ and $g_i$ given by 
\begin{eqnarray}
        \label{ec:loss1}
         g_{i}&=&R_{i}+\gamma Q'(s_{i+1},\mu'(s_{i+1}|\theta^{\mu'})|\theta^{Q'}).
\end{eqnarray}
Here, the discount rate of the predicted future rewards is $\gamma \in [0,1]$. Note that the optimization of the critic network depends on the reward function and the values given by the target networks. The actor is updated with using deterministic policy gradient as
\begin{eqnarray}        \label{ec:actor}        \theta^{\mu}\leftarrow \theta^{\mu}+ \alpha_{a} \frac{1}{N}\sum_{i\in \mathcal{B}} \nabla_{a} Q(s_{i},\widetilde{r}_i|\theta^{Q})\nabla_{\theta^{\mu}}\mu(s_{i}|\theta^{\mu}),
\end{eqnarray}
where the learning rate is $\alpha_a$. The conceptual algorithm that summarizes the DDPG algorithm is shown in the algorithm \ref{DDPGAlgorithm}.

\begin{algorithm}

\caption{DDPG ALGORITHM}
\begin{algorithmic}
\label{DDPGAlgorithm}
\STATE Randomly initialized critic and actor weights $\theta^Q$ and $\theta^{\mu}$
\STATE Initialize target network with $\theta^{Q^{'}} \leftarrow \theta^{Q}$,
   $\theta^{\mu^{'}} \leftarrow \theta^{\mu}$
\STATE Initiate replay buffer
\FOR {episode 1,M}
\STATE Initialize action exploration noise $\mathcal{N}$
\STATE  Receive initial observation state $s_1$
\FOR {episode 1,T}
\STATE Take an action $\widetilde{r}_t$ according to the current policy and add exploration noise $\mathcal{N}$
\STATE Observe the reward $\widetilde{r}_t$ and the new state $s_{t+1}$ 
\STATE Store the transition $(s_t,\widetilde{r}_t,R_t,s_{t+1})$ in replay buffer
\STATE Sample a minibatch of $\mathcal{N}$ transitions from replay buffer
\STATE Update the critic and actor networks according with (\ref{ec:critic})-(\ref{ec:actor})  
\STATE Update the target networks according with (\ref{ec:target})-(3)
\ENDFOR
\ENDFOR
 \end{algorithmic} 
 \end{algorithm}
 
  \section{Human-in-the-Loop: Pitch Angle Control} \label{sec:HIL}
\begin{figure}
\begin{center}
\includegraphics[width=\textwidth]{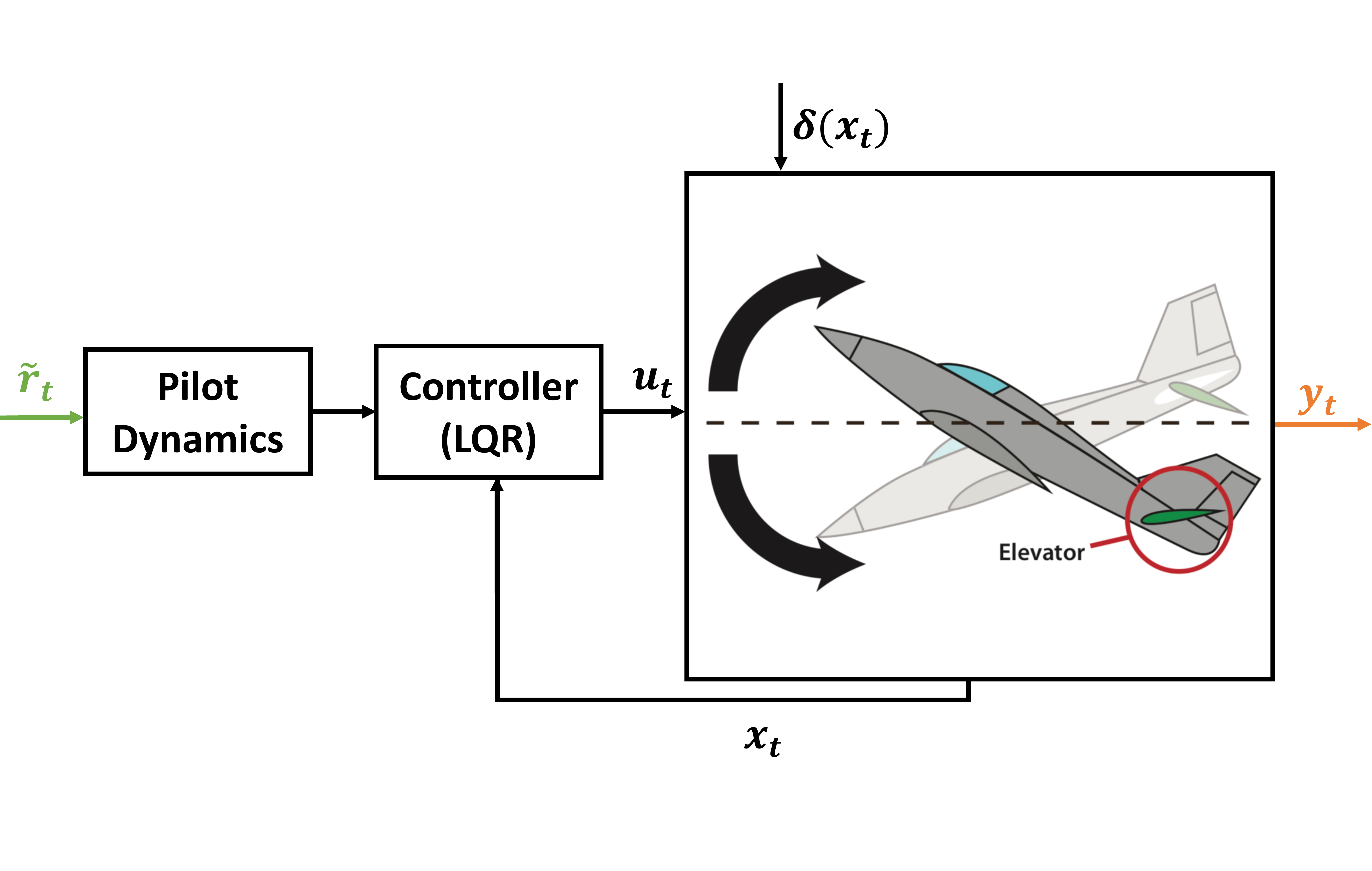}    
\caption{Block diagram of the Human-in-the-Loop System for pitch control for the aircraft Boeing 747.} 
\label{fig:pitch}
\end{center}
\end{figure}

Controlling human in the loop (HIL) systems is a challenging case due to the human reaction delays as a critical parameter for instabilities. We study the pitch angle control problem of an aircraft. This is a problem where the variable to be controlled is the pitch Euler angle (in crad) of the aircraft body axis with respect to the reference axes. The block diagram of the controlled system is illustrated in Fig. \ref{fig:pitch}.

We consider the dynamics given by~\cite{bryson} of longitudinal motion of a Boeing 747 airplane linearized at an attitude of 40kft and velocity of 774 ft/sec. The aircraft is assumed to be operated by a pilot described by a Neal-Schmidt pilot model. In~\cite{tansel}, an online approach is used to guarantee stability and to achieve a desired behavior of the inner-loop system, under the assumption that a model of the system is known a priori. In our DRL-based outer-loop scheme, we assume that the aircraft is controlled by a local linear controller obtained by a LQR approach, and we explore the application of a outer-loop control as a reinforcement learning agent trained using DDPG. 
Let $x_t=[x_{1t}, x_{2t},x_{3t},x_{4t}]^\top$ be the state vector of the dynamical system where $x_{1t}, x_{2t}$ and $x_{3t}$ are the components of the velocity along the three axes (in crad/sec), and $x_{4t}$ represents the pitch Euler angle of the aircraft body axis with respect to the reference axes (in crad). Let $\xi_t\in\mathbb{R}$ be the internal human state vector, and $c_t$ be command produced by the human. A general class of linear human models with constant time delay is given by \cite{tansel}
\begin{align*}
    \label{ec:human_d}
    \dot{\xi}_t&=A_h\xi_t+B_h\theta_{t-\tau} \\
    c_t&=C_h\xi_t+D_h\theta_{t-\tau} \\
    \theta_t&=r_t-x_{4t},
\end{align*}
where $\theta_t$ is the perceived difference between the reference and the pitch angle of the aircraft, $r_t$  is desired reference for the pitch angle. It is assumed that the human has an internal delayed response $\tau\ge0$. This is a linear time-invariant model with time delays known as the Neal-Schmidt model~\cite{NealS} widely used to characterize human dynamics. The dynamics of an aircraft that is controlled by a human are given by
\begin{align*}
        \dot{x}_t&=Ax_t+B(u_t+\delta(x_t))\\
        y_t&=x_{4t}\\
        \dot{x}_{ct}&=x_{4t}-c_t\\
        u_t&=-K_{1}x_t-K_{2}x_{ct}.
\end{align*}
Note that the aircraft's elevator control input $u_t$ is the compound action of a proportion of the state vector $x_t$ and a proportion of the integral of the error between the human command and the pitch angle. Constants $K_1$ and $K_2$ are designed by a LQR design approach. Function $\delta(x_t)\in\mathbb{R}$ models a state-dependent error in the control action. In \cite{tansel}, this expression is defined as $\delta(x_t)=W^T[1, x_{1t}, x_{2t}]^\top$.

\subsection{Simulation Results}
The parameters used to simulate the human-in-the-loop aircraft system are presented in Table \ref{tab:human_value}.
\begin{table}
    \caption{Simulation parameters of the HIL.}
   \label{tab:human_value}
   \centering 
   \begin{tabular}{cc} 
   \hline
    $A_{h}$ & -0.2\\
    $B_{h}$ & 1\\
    $C_{h}$  &  0.08\\
    $D_{h}$  &0.1\\
    $\tau$  &0.5\\
    $A$ & $ \left[ \begin{array}{cccc} -0.003& 0.039& 0& -0.322  \\
    -0.065& -0.319& 7.740& 0  \\
    0.020& -0.101& -0.429& 0  \\
    0& 0& 1& 0 \end{array}\right]$\\
   $B$  & $[0.010, -0.180, -1.160, 0]^\top$\\
   $W$ & $[0.1, 0.3, -0.3]^\top$ \\
   $Q$ & diag([0 0 0 1 2.5]) \\
   $R$ & 10 \\
   \hline
   \end{tabular}
\end{table}
We designed the reward function in Equation \eqref{ec:reward_general} as follows. We define $G(e_t)$ as a linear and decreasing function of the absolute error $G(e_t) = -\alpha_g e_t$, where $\alpha_g=0.3$. In this problem, since $\widetilde{r}_t$ is the reference shown to the human, trajectories with quick changes of the modified reference $\widetilde{r}_t$ are not preferred. Let $\Delta \widetilde{r}_t =\big| \widetilde{r}_t - \widetilde{r}_{t-1}\big|$. Then, we define $H$ as \begin{equation}\label{eq:Hfunction}
H(\widetilde{r}_t)=\begin{cases}
      -\beta_h \Delta \widetilde{r}_t & \textrm{if  }    \Delta \widetilde{r}_t \geq \rho_h  \\
      0 & \textrm{otherwise},
    \end{cases}
\end{equation}
where $\rho_h=3.5$ and $\beta_h=0.1$. Also, since the output of the system is the pitch angle of an aircraft, in this problem, quick changes of the pitch angle variable and values outside a predefined interval are not preferred. Let $\Delta y_t =\big|y_t - y_{t-1}\big|$. Then, $B$ is given by
\begin{equation*}\label{eq:Bfunction}
B(y_t)=\begin{cases}
      -\beta_b \Delta y_t - \delta_b\max (0,\big|y_t\big|-L) & \textrm{if}  \Delta y_t \geq \rho_b   \\
      - \delta_b\max (0,\big|y_t\big|-L) & \textrm{otherwise},
    \end{cases}
\end{equation*}
with $\beta_b=0.1,\delta_b=1, \rho_b=3.5$. This function penalizes quick changes of the output of the system and values outside a boundary defined by $L$. We set $L=18$crad.

\begin{figure}
\begin{center}
\includegraphics[width=0.5\textwidth]{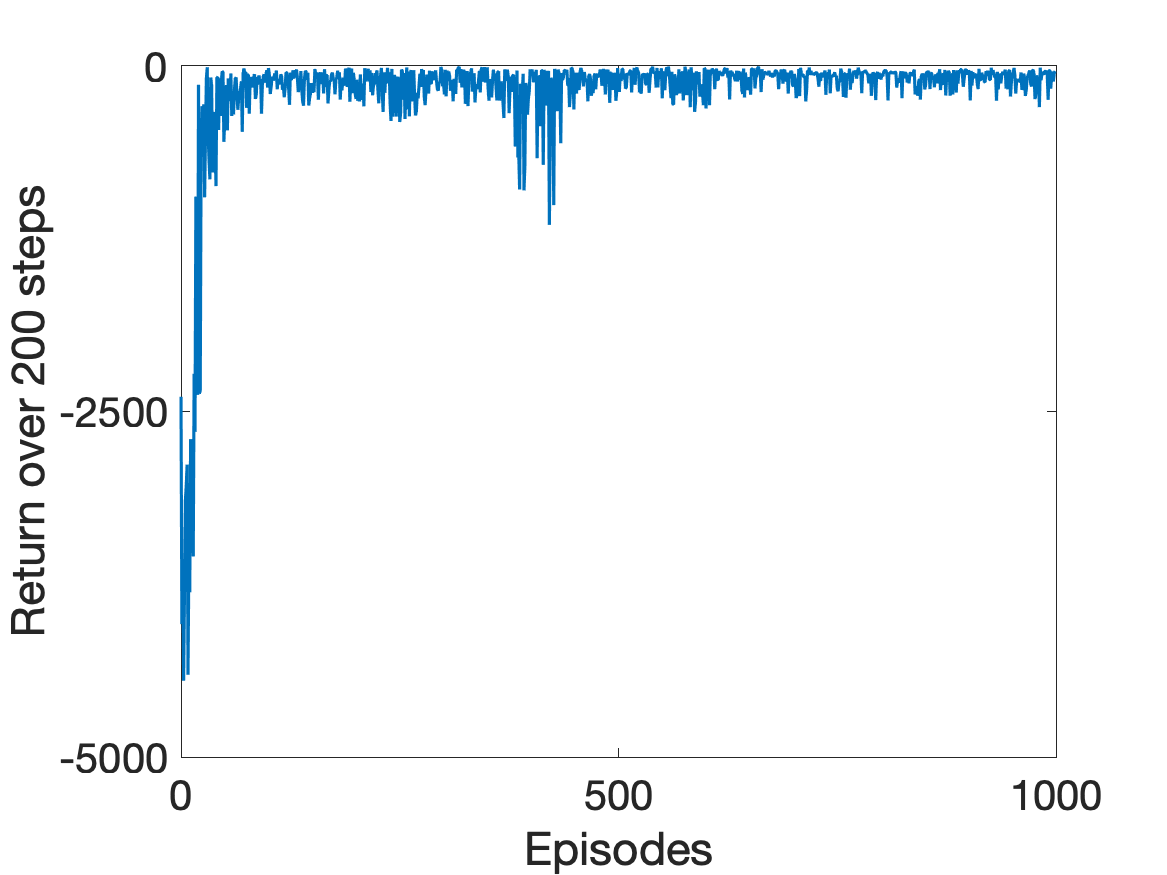}    %
\caption{Average reward per episode in the human in the loop case.} 
\label{fig:recta}
\end{center}
\end{figure}

The learning process is driven by the designed reward function and conducted by randomly choosing an initial condition of the state variables of the system and a reference to be reached. We trained the DRL-based outer-loop control on a 1000 episodes with 200 steps in each episode. Based on the parameters proposed by \cite{Lillicrap}, the learning rates were set to be $\alpha_a=10^{-4}, \alpha_c=10^{-3}$, the discount factor $\gamma=0.99$ and for the soft updates $\tau=0.001$. For the Ornstein-Unlenbeck process parameters were set to $\theta=0.15$ and $\sigma=0.2$. Both the critic and the actor have two hidden layers with 400 and 300 fully connected neurons with a Rectified Linear Units (ReLU) as activation functions. The output layer the actor was a tanh function to bound the actions. Although the critic takes the state and action as inputs, the action is included in the second layer. Finally, we used ADAM \cite{adam} as a optimization method for learning the neural network parameters. This is an adaptive learning rate optimization algorithm designed to train deep neural networks. Fig. \ref{fig:recta} shows the performance of the training process over 1000 episodes with 200 steps. Note that a finite time convergence in the average reward is presented. At the beginning of the training process the average reward per episode is poor due to the exploration stage of the system.  After 500 episodes the RL agent was able to learn a consistent optimal policy. 

\begin{figure}
    \centering
    \begin{subfigure}[b]{0.45\textwidth}
        \includegraphics[width=\textwidth]{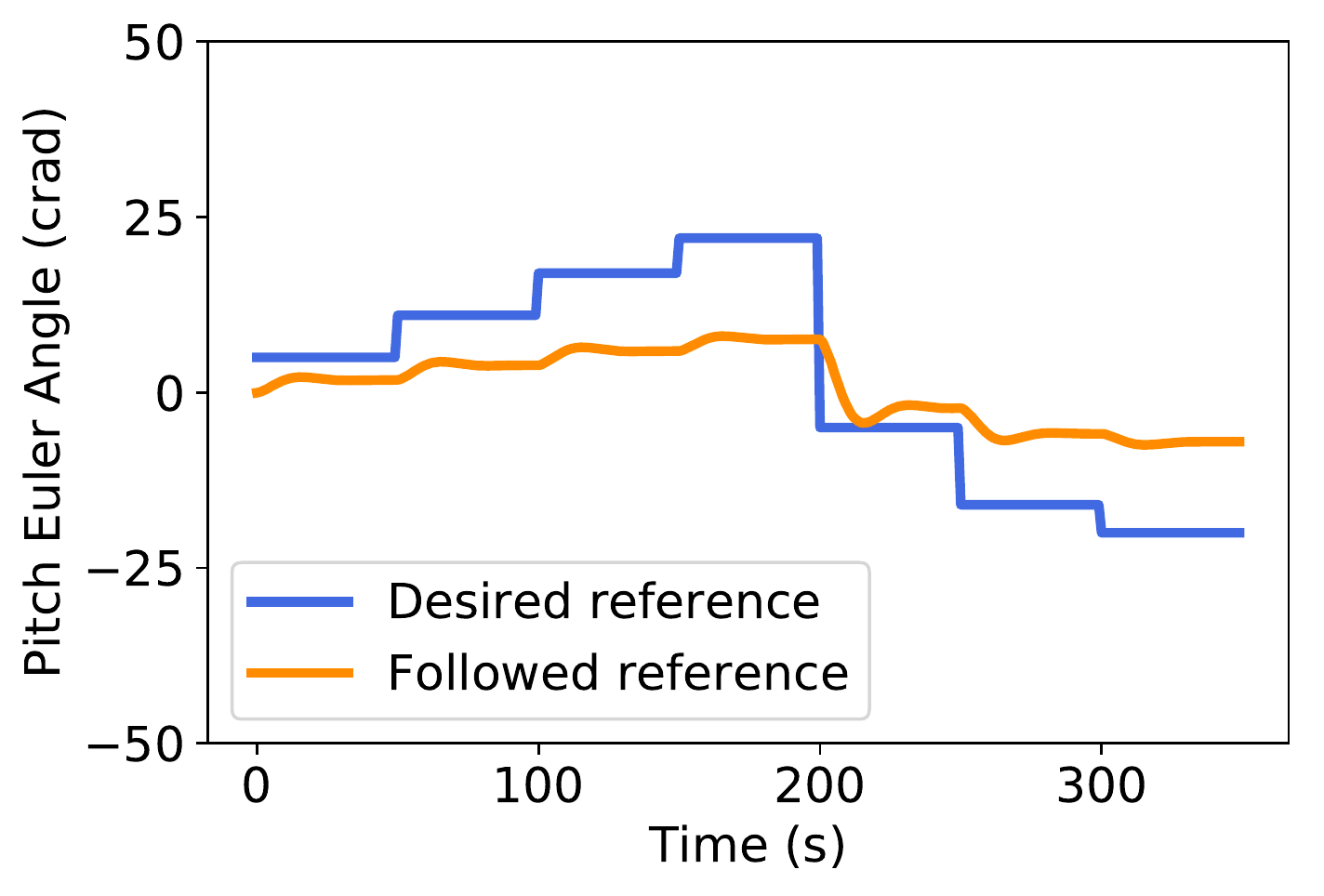}
        \caption{}
        \label{fig:pitch_HIL_h}
    \end{subfigure}
    ~ 
    \begin{subfigure}[b]{0.45\textwidth}
        \includegraphics[width=\textwidth]{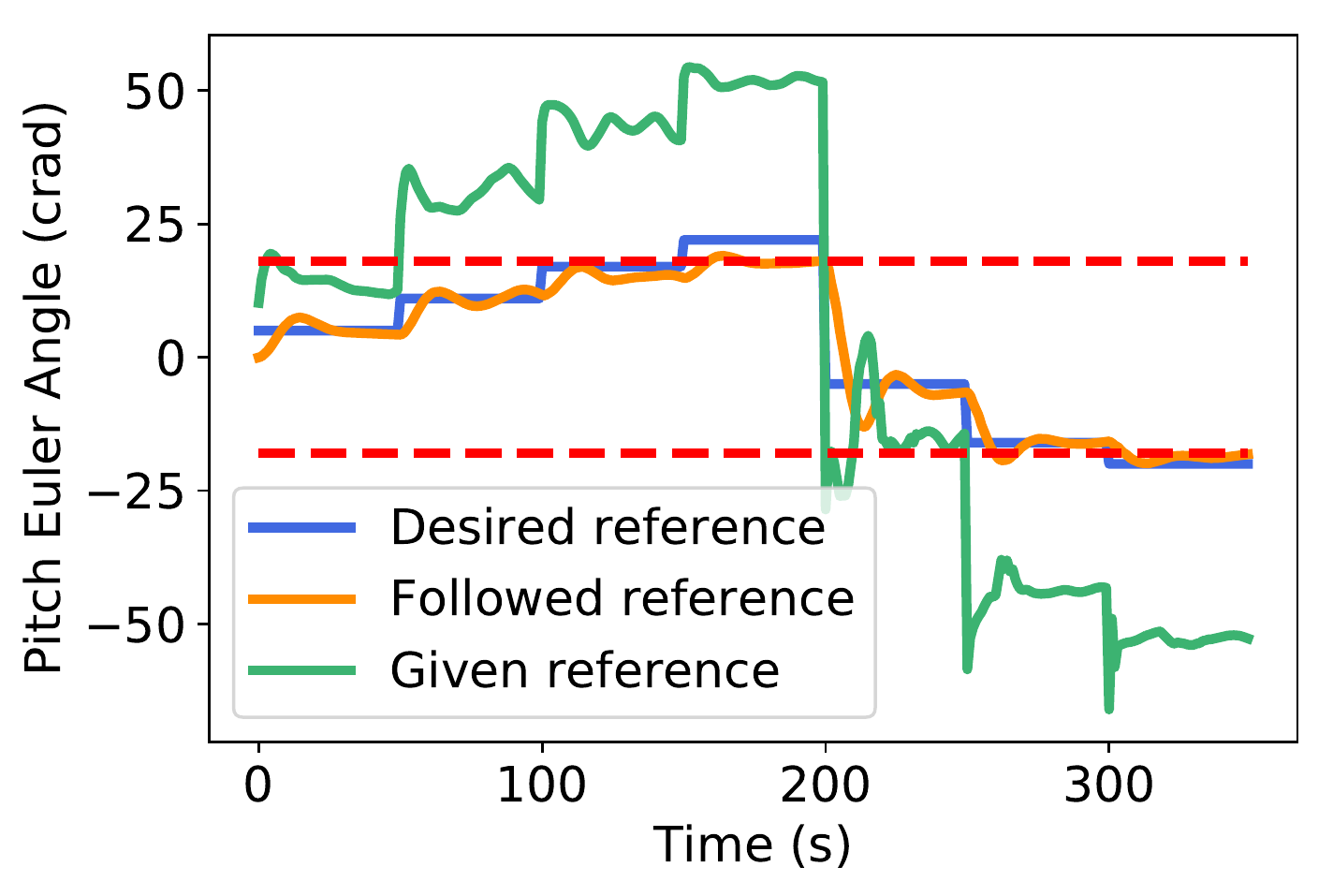}
        \caption{}
        \label{fig:pitch_DRL_h}
    \end{subfigure}
    ~ 
    \begin{subfigure}[b]{0.45\textwidth}
        \includegraphics[width=\textwidth]{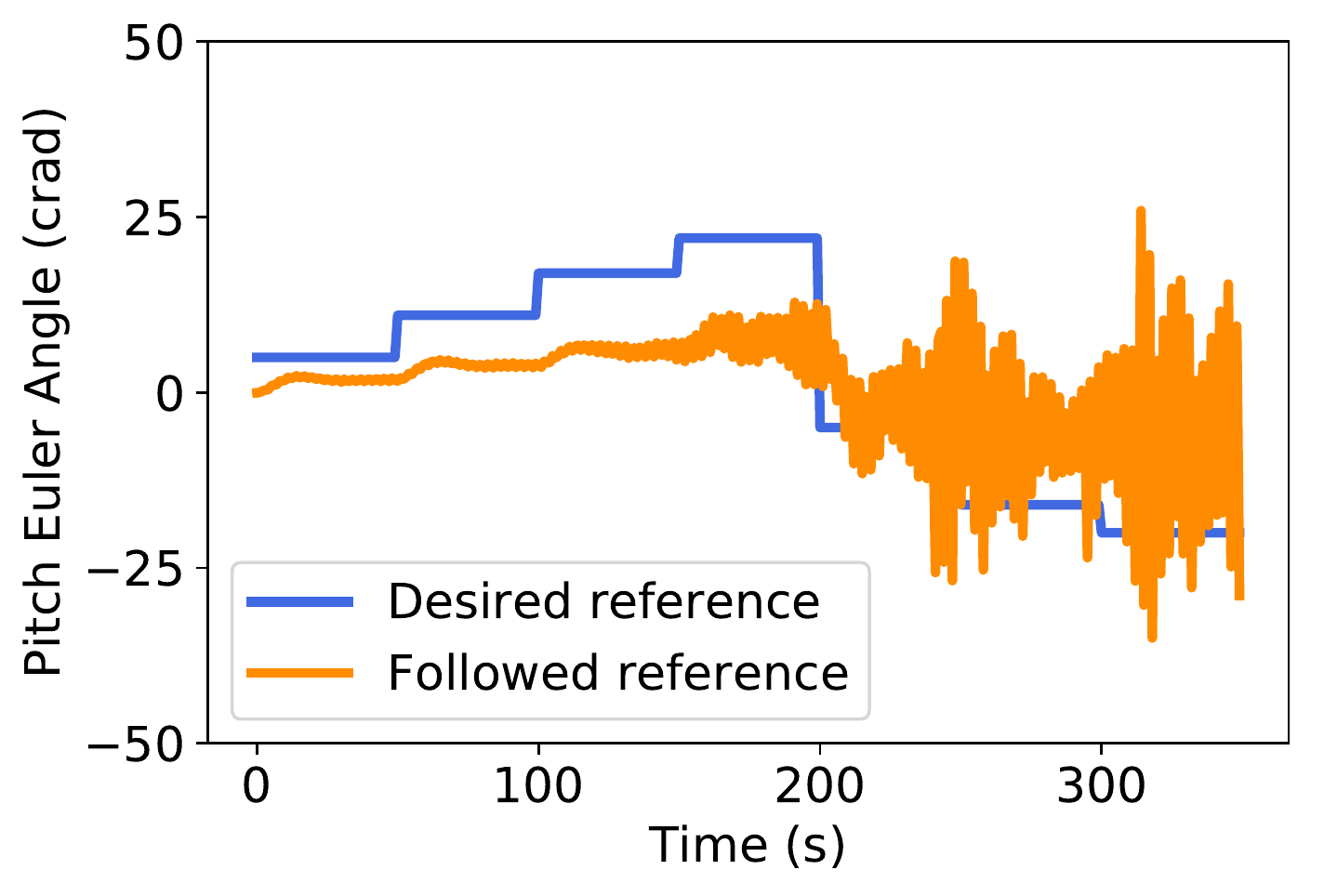}
        \caption{}
        \label{fig:pitch_HIL}
    \end{subfigure}
    \begin{subfigure}[b]{0.45\textwidth}
        \includegraphics[width=\textwidth]{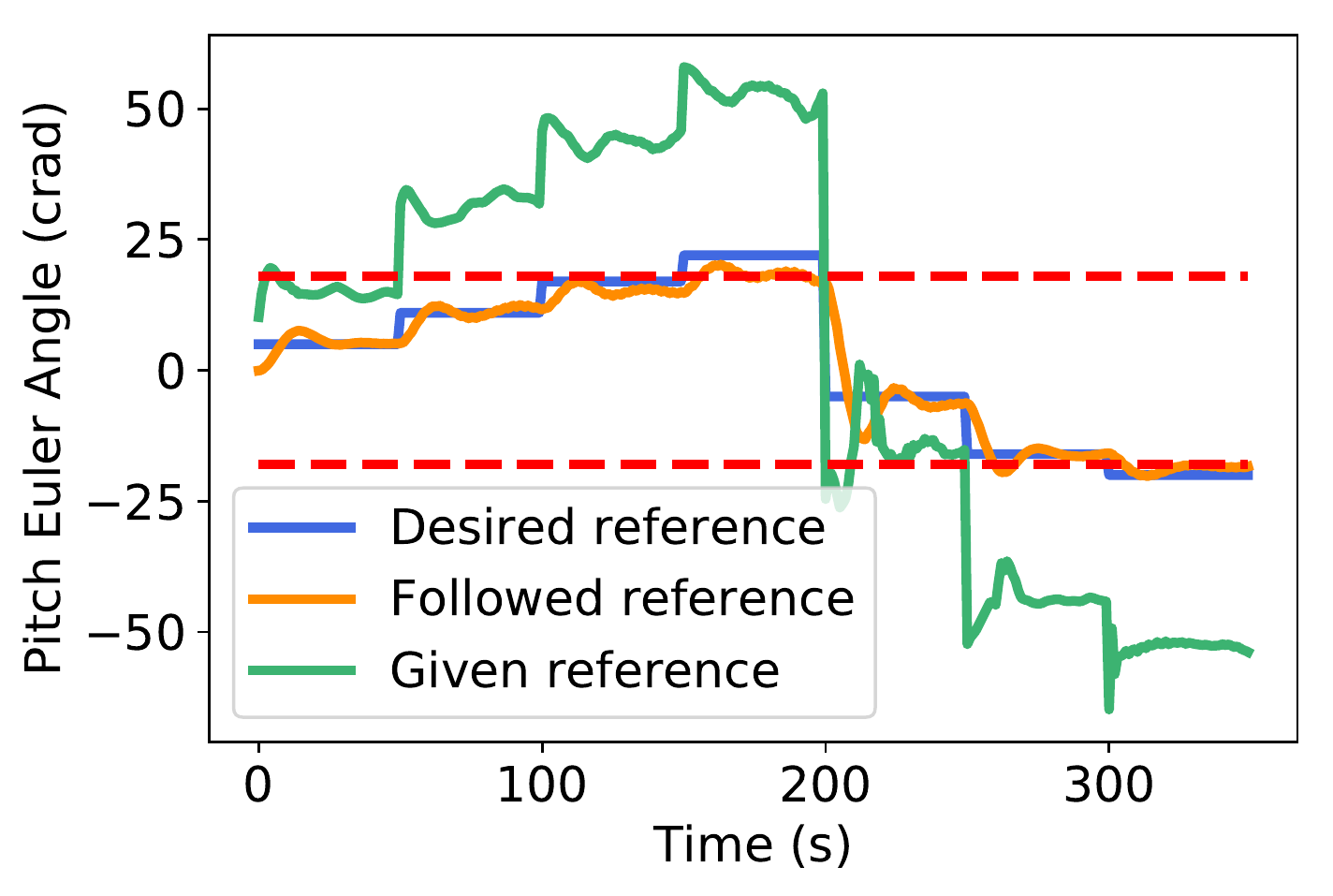}
        \caption{}
        \label{fig:pitch_DRL}
    \end{subfigure}
    \caption{Human in the Loop - Pitch control. (a) Inner-loop system response with mean absolute error (MAE of 10.91crad). (b) Inner-loop system response with DRL-based outer-loop control (MAE of 3.90crad). (c) Inner-loop system response with instabilities (MAE of 10.13crad). (d) Inner-loop system response with instabilities and DRL-based outer-loop control (MAE of 2.46crad).}\label{fig:hil_control}
\end{figure}

After the training stage, we tested the DRL-based outer-loop control in two trajectory tracking scenarios: an aircraft in normal operation and an aircraft with a local control action based on delayed information. Fig. \ref{fig:pitch_HIL_h} shows the inner-loop response without the trajectory tracking control. Fig. \ref{fig:pitch_DRL_h} shows the system response to the modified reference given by the DRL-based trajectory tracking. Note that the proposed outer-loop control was able to find a policy that tries to satisfy the system's constraints while minimizing the tracking error. The mean absolute error (MAE) for the system without the outer-loop control is 10.91crad and with the outer-looping scheme it is 3.90crad. This is an improvement of $64.25\%$ in the MAE. Note that the output of the system in the trajectory tracking scheme remains inside the defined boundaries.

We also tested the performance of our learned policy in a scenario where there is a delay of $5.7s$ in the controller of the aircraft, that is, $u_t=-K_1x_{t-5.7}-K_2x_{c(t-0.57)}$. Fig. \ref{fig:pitch_HIL} shows the inner-loop response with instabilities due to the delays in the controller, and Fig. \ref{fig:pitch_DRL} shows the system response to the given reference. The DRL-based outer-loop control was able to minimize the tracking error, to enforce systems constraints, and to provide a smooth output signal in a scenario that was not directly part of the learning process. The mean absolute error (MAE) for the system without the outer-loop control is 10.13crad, and 2.46crad with the outer-loop scheme. This is an improvement of $75.71\%$.
 
 \section{Mean Field Games: Electric Space Heaters} \label{sec:MFC}
\begin{figure}
\begin{center}
\includegraphics[width=\textwidth]{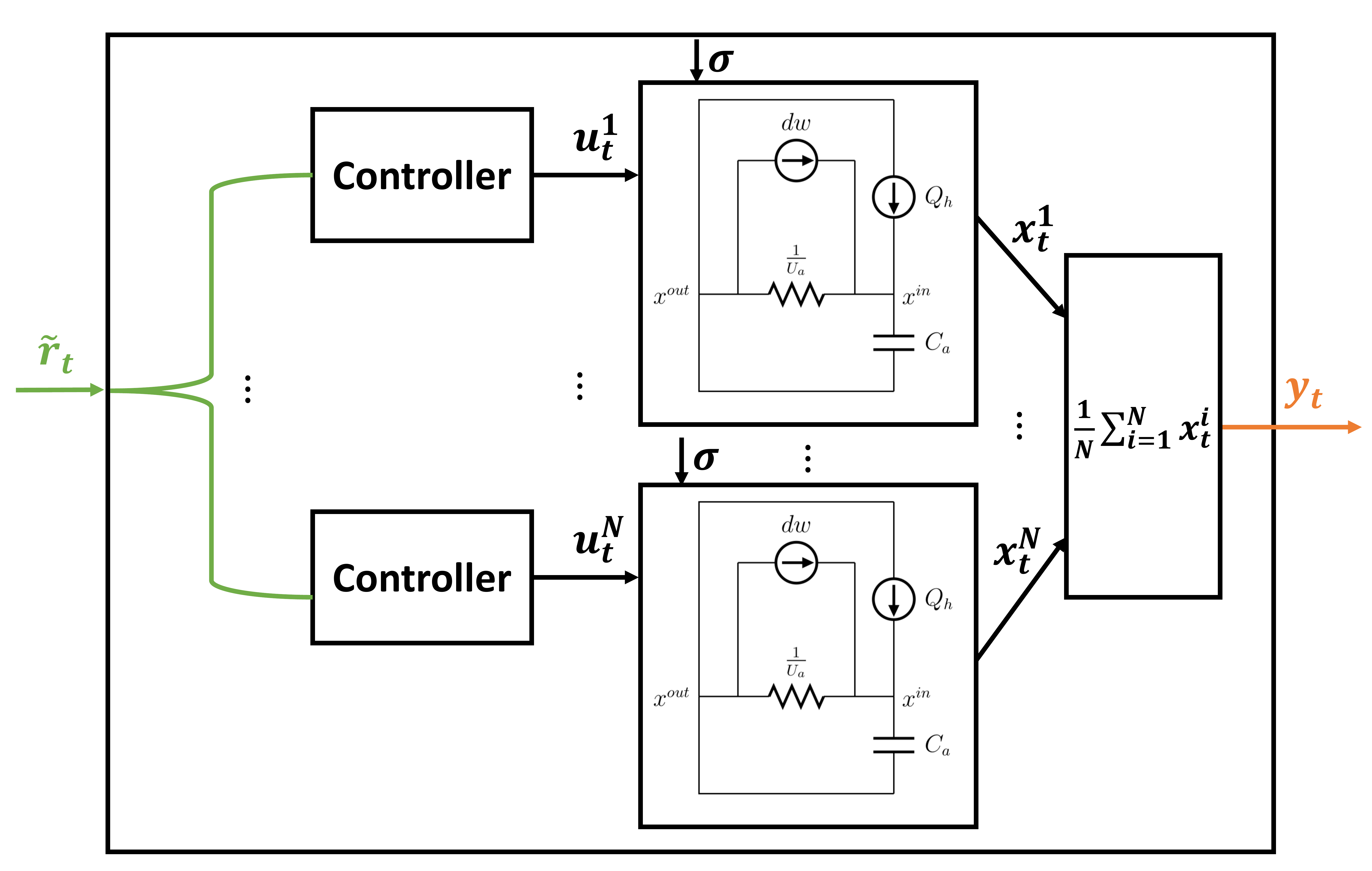}    
\caption{Block diagram of the massive number of controlled space heaters.} 
\label{fig:circuitHeaters}
\end{center}
\end{figure}

The second study case involves mean field games (MFGs) and populations with many individuals that have common objectives. MFGs are commonly used to define strategies to control large-scale complex multi-agent systems \cite{malhame,caines}. We study a power system problem where it is required to control the average temperature of a massive number of space heating devices in a way it follows a target temperature trajectory \cite{malhame}. The block diagram of the controlled system is illustrated in Fig. \ref{fig:circuitHeaters}.

In \cite{caines}, the dynamics of the space heating devices are described by a stochastic linear system, and this problem is solved with a linear control law based on MFGs under the assumption that the agents are dynamically independent and cost-coupled. Even though the linear control law provides some theoretical guarantees and implementation benefits, its ability to lead the system to follow the desired reference needs to be improved. Our motivation to improve the performance of the system with the linear control law is the infeasibility and cost inefficiency of modifying the local controller of every household electric space heater. We show how our DRL-based outer-loop scheme is able to improve the performance of the system in a model-free fashion without modifying the local linear control based on MFGs.

Let ${x}^{i}_t\in\mathbb{R}$ be the air temperature inside the household $i$ at time $t$. The thermal dynamics of the household that has a space heater $i$ are modeled using the stochastic differential equation \cite{malhame}
\begin{align}
\label{ec:dynamic_heater}        
        d{x_t}^{i}&=\frac{1}{C_{a}}[-U_{a}({x_{t}}^{i}-{x}^{out})+Q_{h}]dt+\sigma d\omega_t^i \nonumber\\
         Q_h&={u_{t}}^{i}+u^{free}\\
        u^{free} &\triangleq \frac{-U_a}{C_a}(x^{out}-{x_0}^{i}),\nonumber
\end{align}
where ${x}^{out}$ is the outside ambient temperature, ${x_{0}}^{i}$ is the initial temperature of each household, $C_{a}$  is the thermal mass of air inside, $U_{a}$ is the conductance of air mass inside, $Q_{h}$ is the heat flux from the heater, $w_{t}$ is a standard Wiener process that characterizes perturbations on the system, $\sigma$ is a volatility term, and $N$ is the number of individuals. The heat flux from the heater depends on the control action $u_t^i$. Here, the goal of the local linear control strategy is to make $x_i$ reach a reference given by a command center. This reference is the same one for all space heaters. The action of a controlled space heater $i$ is such that the heater tries to follow the mass of space heaters and the energy of the control signal $u_i$ is minimized. This behavior is modeled 
as the minimization of the cost function
$$J_i(u_i)\triangleq E \left[ \int_0^\infty {e^{-\phi t} \left[ ({x}^{i}_t-{v}^{i}_t)^2+r ({u}^{i}_t)^2 \right] dt }\right],$$
subjected to the dynamics of the system modeled by the stochastic differential equation in \eqref{ec:dynamic_heater}. Signal ${v}^{i}$ measures the average effect of the mass formed by all the other individuals:
$${v}^{i}_t=\gamma \left(\frac{1}{n}\sum_{k \neq i }^n {x}^{k}_t +\eta \right).$$ 

According to \cite{caines}, for a very large population of individuals, one can assume that $v_t^i\approx x_{t}^{*}$ for all $i$, where $x^{*}$ is a deterministic function given by  $x_{t}^{*}=\gamma (\frac{1}{n} \sum_{i=1}^n E[x_t^i]+ \eta)$. The main goal of the control action $u_t^i$ is to drive household $i$ to minimize its own cost function while the mean field of the population reaches a reference temperature $x_{\infty}^*$ at steady state. The linear control law that each household locally implements is given by
\begin{align*}
        {u_{t}}^{i}&=-\frac{1}{C_ar}(\Pi_a{x_{t}}^{i}+{s_{t}}^{i}) \\
        {s_{t}}^{i}&=s_{\infty}+\left(\frac{\gamma}{\beta_2-\lambda_1}\right)({x}_{\infty}^{*}-{x_0}^{i})e^{\lambda_1t}\\
        s_{\infty}&=\frac{{x}_{\infty}^{*}}{\frac{-U_{a}}{C_{a}}-\frac{1}{C_{a}^{2}r}\Pi_{a}-\phi}, 
\end{align*} 
where $\Pi_a$ is a constant that comes from solving an algebraic Riccati equation, $s_t$ is an offset of the control action, and $x_0^i$ is the state at $t=0$. The interpretation of the remaining parameter $\lambda_1$ is given in detail in \cite{caines}. The incorporation of this controlled system into the DRL-based outer-loop control scheme in Fig. \ref{bg} involves the application of a modified reference such that the mean temperature defined as $y_t=\frac{1}{N}{\sum_i} {x_t}^{i}$ accurately follows the desired reference. Details on the parameter design can be found in \cite{caines}.

\subsection{Simulation Results}
The simulation parameters of the space heaters are set as $C_{a}= 10kWh/^{\circ}C$, $U_{a}= 0.2kW/^{\circ}C$, $x^{out}=-5^{\circ}C$,    $\sigma=0.25$, $\Pi_{a}=0.4$, $r=10$,    $\delta=0.001$, $\gamma=0.6$, $\eta=0.25$, $\beta_{1}=0.038$, and $\lambda_{1}=-0.028$. The reward function is defined as follows. $G(e_t)$ in Equation  \eqref{ec:reward_general} is given by $G(e_t) = -\alpha_g e_t$, where $\alpha_g=0.3$. Function $G(e_t)$ is decreasing and linear with slope $-\alpha_g$. Since it is not convenient to force the system to have sudden changes in the modified reference shown to each electric space heater, function $H$ is defined as in Equation \eqref{eq:Hfunction} with $\beta_h=0.1$ and $\rho_h=1$. In this problem, we consider a comfort constraint that dictates that the maximum temperature change per hour on each household is $1^{\circ}C$. To enforce the comfort constraint, $B$ is defined as
\begin{equation*}\label{eq:Bfunction2}
B(y_t)=\begin{cases}
      -\beta_b \Delta y_t - \delta_b\max (0,\Delta y_t-L) & \textrm{if}  \Delta y_t \geq \rho_b   \\
      - \delta_b\max (0,\Delta y_t-L) & \textrm{otherwise},
    \end{cases}
\end{equation*}
with $\rho_b=\delta_b=L=1$, $\beta_b=0.1$ and $\Delta y_t=|y_t-y_{t-1}|$.
This reward function significantly improves tracking error and enforces the comfort constrain to be satisfied.

\begin{figure}[!ht]
\begin{center}
\includegraphics[width=0.5\textwidth]{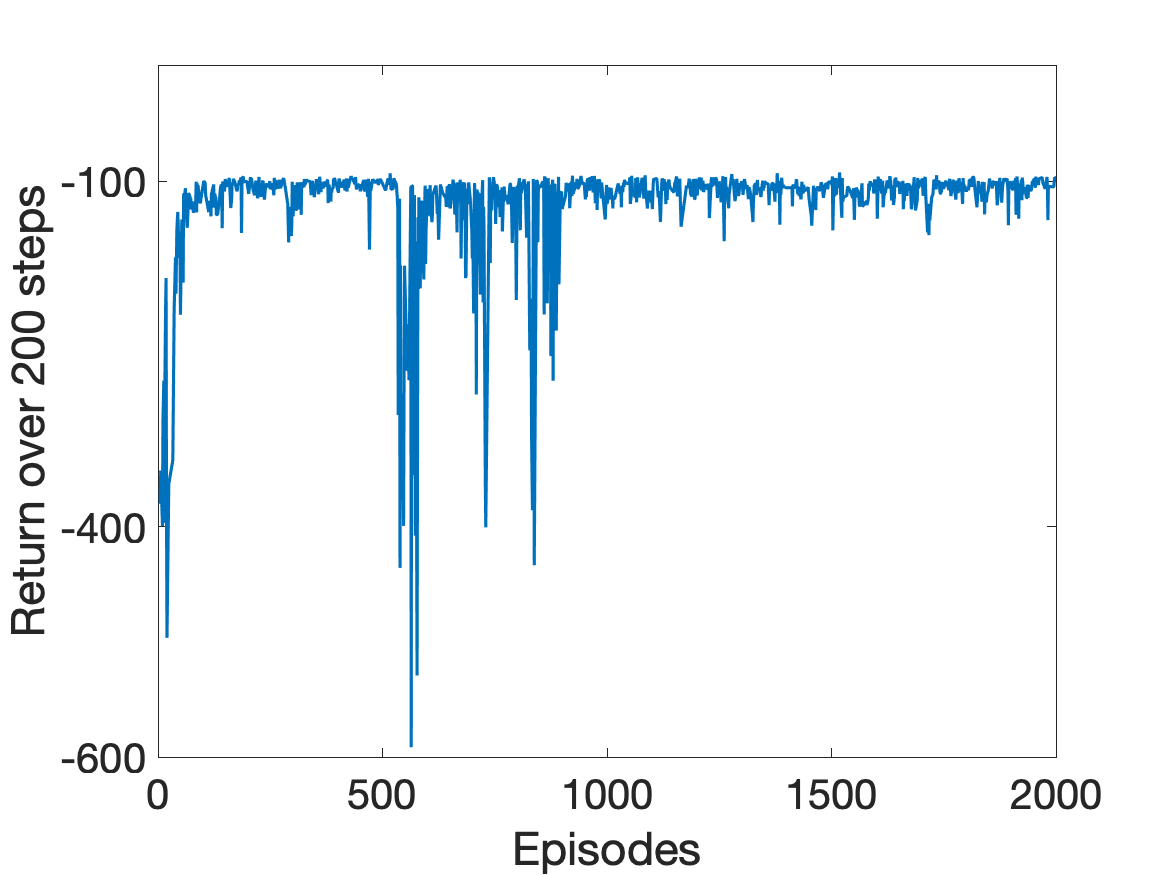}    
\caption{Average reward per episode in the mean field games case.} 
\label{fig:mfg}
\end{center}
\end{figure}

The learning process is conducted by randomly choosing an initial condition of the state variables of the system and a reference to be reached for four electric space heaters. We trained the outer-loop control on a 2000 episodes with 200 steps in each episode. Based on the parameters proposed by \cite{Lillicrap}, the learning rates were set to be $\alpha_a=10^{-4}, \alpha_c=10^{-3}$, the discount factor $\gamma=0.99$ and for the soft updates $\tau=0.001$. For the Ornstein-Unlenbeck process the parameters were $\theta=0.15$ and $\sigma=0.2$. Both the critic and the actor have two hidden layers with 400 and 300 fully connected neurons with a Rectified Linear Units (RELU) as  activation function. The output layer the actor was a tanh, to bound the actions. Although the critic takes the state and action as a inputs, the action is include in the second layer. Finally, for learning the neural network parameters we use ADAM \cite{adam} as the optimization method. Fig. \ref{fig:mfg} shows the performance of the training process over 2000 episodes with 200 steps. Note that there is a finite time convergence in the average reward. At the beginning of the training process, the average reward per episode is poor since the system is in an exploring stage. After 1000 episodes, the RL agent was able to learn a consistent optimal policy. 

\begin{figure}
    \centering
    \begin{subfigure}[b]{0.45\textwidth}
        \includegraphics[width=\textwidth]{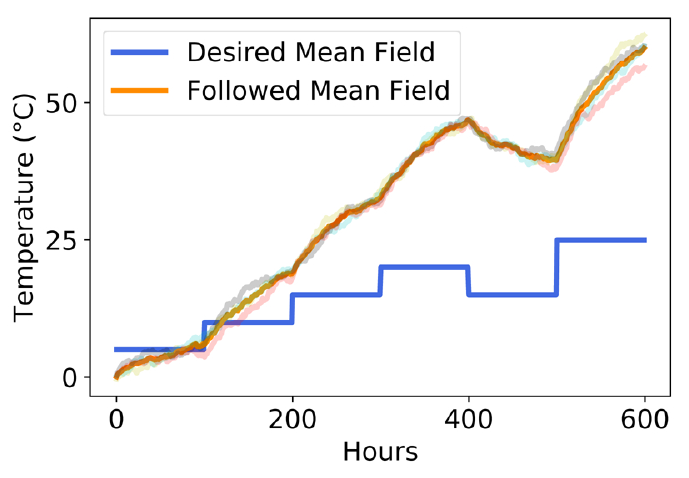}
        \caption{}
        \label{fig:mfg_org}
    \end{subfigure}
    ~ 
    \begin{subfigure}[b]{0.45\textwidth}
        \includegraphics[width=\textwidth]{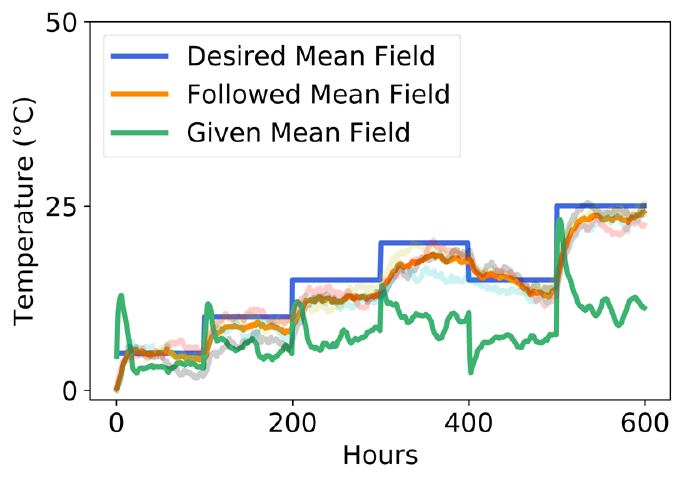}
        \caption{}
        \label{fig:mfg_RL}
    \end{subfigure}
    ~ 
    \begin{subfigure}[b]{0.45\textwidth}
        \includegraphics[width=\textwidth]{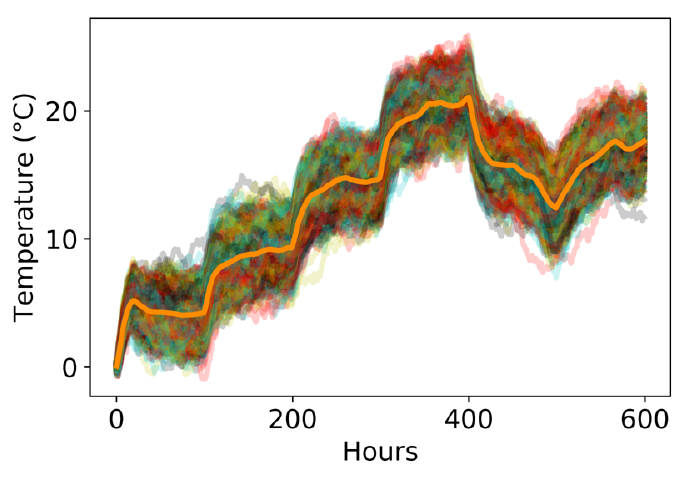}
        \caption{}
        \label{fig:t_heater}
    \end{subfigure}
    \begin{subfigure}[b]{0.45\textwidth}
        \includegraphics[width=\textwidth]{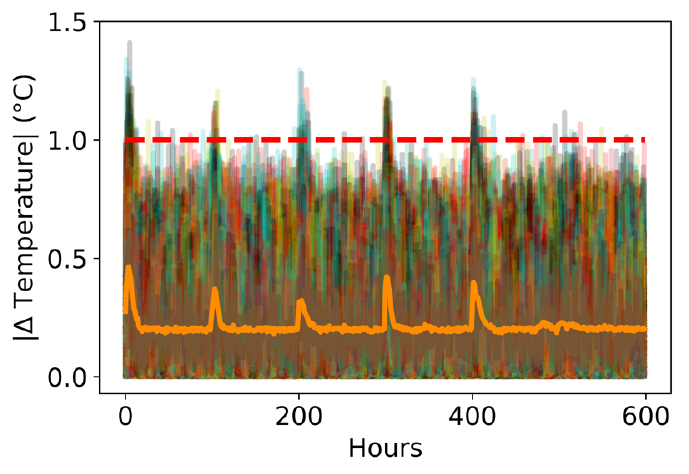}
        \caption{}
        \label{fig:mfg_constr}
    \end{subfigure}
    \caption{Mean Field Control - Electric Space Heaters. (a) Inner-loop system response with MAE of $15.62^{\circ}C$. (b) Inner-loop system response with DRL-based outer-loop control with MAE of $1.17^{\circ}C$. (c) Inner-loop system response with DRL-based outer-loop control for 1000 space heaters with MAE improvement of $91.3\%$. (d) Absolute temperature  difference for 1000 heaters.}\label{fig:mfg_control}
\end{figure}

After training stage, we tested the DRL-based outer-loop control in two trajectory tracking scenarios: a mean field problem with 4 electric space heaters and a mean field problem with 1000 heaters. The simulation results of the mean field tracking problem with no outer-loop control is shown in Fig.~\ref{fig:mfg_org}. As predicted, the linear control action leads the space heaters to an underperforming trajectory. Fig.~\ref{fig:mfg_RL} shows that the outer-loop control is able to learn an optimal policy that allows the system to achieve the desired behavior in a model-free fashion while preserving the linear control law. The MAE of the MFG system with no outer-loop control was $15.62^{\circ}C$, and for the system under the outer-loop control scheme was $1.17^{\circ}C$, which is an improvement of $92.5\%$ in the MAE. Additionally, using the policy learned by the DRL-based outer-loop control from the scenario with 4 electric space heaters, we tested a system with 1000 space heaters. In this case, the MAE was improved from $15.07^{\circ}C$ with just the local controller to $1.31^{\circ}C$ when our outer-loop control was added. The results are shown in Fig.~\ref{fig:t_heater}. In Fig.~\ref{fig:mfg_constr}, we present the temperature change per hour to show that the comfort constraint is satisfied for the mean field. However, some in some households this constraint is violated when the desired reference changes abruptly and the system focuses on minimizing the tracking error as soon as possible. It is important to note that a reward without the $B$ function would lead to a higher violation of the constraint in order to achieve a faster tracking task. 
 
\section{Conclusion}
\label{sec:conclusion}
We presented an exploratory study of a data-driven and non-invasive outer-loop control using deep reinforcement learning for trajectory tracking. The main objective of this DRL-based outer-loop control was to provide a modified reference to the trajectory-tracking system such that the performance of the system is improved. The DRL-based outer-loop control was seen as an agent whose learning process is driven by a reward function that depends only on the output of the controlled system and the reference to be reached by the trajectory-tracking system. We proposed a general form of the reward function that leads the system to minimize the tracking error and that penalizes state values and modified reference values that do not satisfy previously defined constraints. 
We analyzed two challenging study cases where the overall performance of the system was significantly improved: a human in the loop system with time delays, and a control system involving a massive number of electric space heaters where the local control is ruled by a mean field game-based control action. We showed that the DRL-based outer-loop control is able to improve the performance of the controlled systems.\\
Future directions include study strategies to have a safe exploration during the learning process to ensure critical constraint satisfaction \cite{safeRL}. Also, recent theoretical analyses of control techniques based on safe reinforcement learning can be applied to study properties of the DRL-based outer-loop control for trajectory tracking systems \cite{lyapunov}. 

\bibliographystyle{unsrt}  


\bibliography{paper}






\end{document}